\documentclass{ws-procs9x6}

\def\etal {{\it et al.}}
\usepackage{multicol}

\begin{document}

\title{SENSITIVITY OF ATMOSPHERIC NEUTRINOS IN SUPER-KAMIOKANDE 
TO LORENTZ VIOLATION}

\author{T.\ AKIRI}

\address{Physics Department, Duke University\\
Durham, NC, 27705, USA\\
E-mail: tarek.akiri@duke.edu}

\author{On behalf of the Super-Kamiokande Collaboration}

\begin{abstract}
This talk, given at CPT'13, showed Super-Kamiokande atmospheric-neutrino Monte Carlo sensitivity to Lorentz-violation effects using the perturbative model derived from the Standard-Model Extension.
\end{abstract}

\bodymatter

\section{Introduction}

The Standard-Model Extension~\cite{SME} (SME) is an effective field theory having all the features of the Standard Model but adding all possible Lorentz-violating terms. Recently, many experiments have been using this framework to test Lorentz invariance. Neutrino oscillations, as an interferometric effect, are a very sensitive probe for Lorentz violation (LV) effects expected to manifest around the Planck scale.
Many neutrino oscillation experiments used either the short-baseline approximation or the perturbative model to search for sidereal variations in their data constraining the neutrino LV coefficients~\cite{NuLV}.

Super-Kamiokande~\cite{SKNIM} (SK) is an underground 50 kT water Cherenkov detector located in Kamioka (Japan).   Its innermost volume is instrumented with 11146 20'' photomultiplier tubes (PMTs) that allow the reconstruction of neutrino interaction features based on the time and charge of the hit PMTs. In 1998, the analysis of SK atmospheric-neutrino data proved the neutrino oscillation phenomenon~\cite{SK1998} through the disappearance of $\nu_\mu / \bar{\nu}_\mu$ and the non-appearance of $\nu_e / \bar{\nu}_e$. 
Using SK atmospheric-neutrino Monte Carlo (MC) and the SME perturbative model~\cite{PertLV}, we performed a sensitivity study for isotropic Lorentz-violation effects.

\section{The perturbative model}

The perturbative model is derived from the SME using time-dependent perturbation theory. 
The LV Hamiltonian is derived up to second order in the perturbative series for both $\nu \rightarrow\nu$ and $\nu\rightarrow\bar\nu$ oscillations. In this study, we restricted ourselves to $\nu\rightarrow\nu$ oscillations leaving two effective sets of coefficients: 
$a_{\rm eff}$ (CPT-odd) and $c_{\rm eff}$ (CPT-even), henceforth denoted as $a$ and $c$, respectively. 
Furthermore, we choose to consider only the isotropic and renormalizable part that leads to spectral distortions. The perturbative Hamiltonian has then the following form:
\[ \delta h = \frac{1}{|p|}\left( \begin{array}{ccc}
a_{ee}-c_{ee}\ \ \ \                  & a_{e\mu}-c_{e\mu}\ \ \ \                    & a_{e\tau}-c_{e\tau} \\
a_{e\mu}^*-c_{e\mu}^*\ \ \ \  & a_{\mu\mu}-c_{\mu\mu}\ \ \ \           & a_{\mu\tau}-c_{\mu\tau}  \\
a_{e\tau}^*-c_{e\tau}^*\ \ \ \  & a_{\mu\tau}^*-c_{\mu\tau}^*\ \ \ \    & a_{\tau\tau}-c_{\tau\tau} 
\end{array} \right). \]
%
\label{sec:LVPheno}
LV effects can be easily described by considering $\nu_\mu$ disappearance at the first order in the two-flavor case:
\begin{equation}
P_{LV} (\nu_\mu\rightarrow \nu_\mu) = \sin(2.534\times\frac{\Delta m^2 L}{E})\times (\Re e (c_{\mu\tau})LE - \Re e (a_{\mu\tau})L),
\label{eq:2DProb}
\end{equation}
with $\Delta m^2$ being the atmospheric mass splitting in eV$^2$, $L$ the neutrino pathlength in km and $E$ its energy in GeV. The LV coefficients $a_{\mu\tau}$ expressed in km$^{-1}$ and $c_{\mu\tau}$ in km$^{-1}$GeV$^{-1}$ are complex.

\begin{figure}
\psfig{file=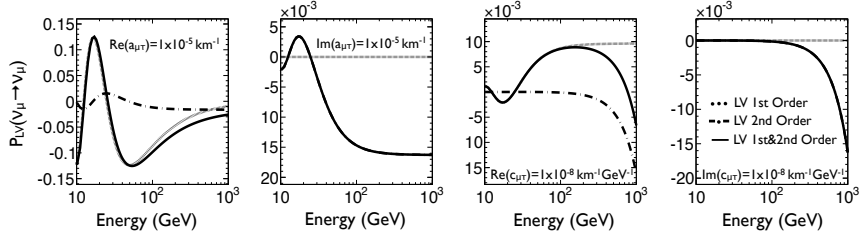,width=\textwidth}
\caption{Illustration of the first (dotted line), second (dashed line) and first plus second order (solid line) LV $\nu_\mu\rightarrow\nu_\mu$ oscillation probabilities for the four $\mu\tau$ coefficients taken individually as a function of energy for $L\simeq$12800 km.}
\label{fig:LVProbs}
\end{figure}

Equation~\eqref{eq:2DProb} shows that $a_{\mu\tau}$ and $c_{\mu\tau}$ control oscillations proportional to $L$ and $L \times E$ respectively, each with opposite signs. In the two-flavor case, only the real parts of the LV coefficients are involved. 
Calculating the probabilities in the three-flavor case implies the imaginary parts as well.
For a given value of the LV coefficients at first order in the perturbative series, the imaginary parts gives much smaller probabilities than the real parts. 
In our analysis, we extended the LV calculation up to the second order since the latter is expected to be the dominant contribution in the no-oscillation region corresponding to high energy in SK.
Indeed, looking at the sine term in Eq.\ \eqref{eq:2DProb}, one can see that at high energy the first order probability is suppressed in contrast to the second order (see Fig.\ \ref{fig:LVProbs}). Moreover, the probability for the real and imaginary parts is similar for the second order. 

\begin{figure}
\psfig{file=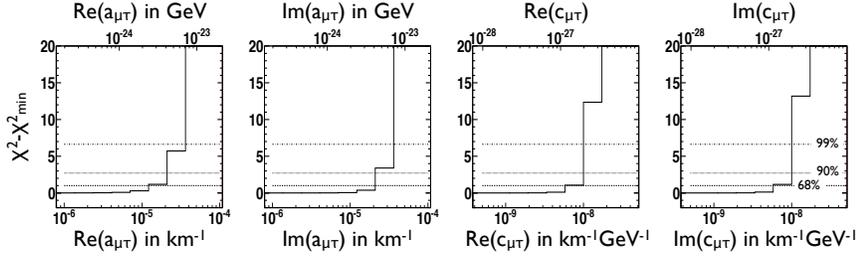,width=\textwidth}
\caption{$\Delta \chi^2$ as a function of the LV coefficient value tested along with the 68\%, 90\% and 99\% C.L. lines. From left to right: Re($a_{\mu\tau}$), Im($a_{\mu\tau}$), Re($c_{\mu\tau}$) and Im($c_{\mu\tau}$).}
\label{fig:LVSens}
\end{figure}

\section{Analysis}
In its three-flavor oscillation analysis, SK divides its data into many different categories based on the event topology and the event reconstructed features~\cite{SK3flavors}. 
The cosine of the zenith angle, related to the neutrino pathlength, and the energy are used to further divide each category in zenith angles and momentum bins for a total of 480 analysis bins. SK has many systematic uncertainties arising mainly from the neutrino flux and interaction as well as from detection effects. In this study, except for the CP phase $\delta$, we also considered systematic errors associated with oscillation parameters: $\Delta m^2_{21}=7.46\times 10^{-5} \text{ eV}^2, \Delta m^2_{32}=~2.44\times 10^{-3} \text{ eV}^2, \sin^2\theta_{12}=0.32, \sin^2\theta_{23}=0.5 \text{ and } \sin^2\theta_{13}=0.0251$ for a total of 159 systematic errors. $\delta$ has been taken as a parameter because it causes both the period and amplitude of the oscillations to change, even switching sign. We use a poissonian $\chi^2$ that is minimized iteratively at each LV coefficient value tested. The sensitivity for each of the LV coefficients tested individually is shown in Fig.\ \ref{fig:LVSens}.
The sensitivity obtained at 90\%~C.L. is:
\begin{multicols}{2}
\begin{itemize}
\item $\text{Re}(a_{\mu\tau}) < 4.1 \times 10^{-24} \text{ GeV}$,
\item $\text{Im}(a_{\mu\tau}) < 5.1 \times 10^{-24} \text{ GeV}$,
\item $\text{Re}(c_{\mu\tau}) < 1.7 \times 10^{-27}$,
\item $\text{Im}(c_{\mu\tau}) < 1.7 \times 10^{-27}$.
\end{itemize}
\end{multicols}

Note that the sensitivity is comparable for the real and imaginary parts of both coefficients. 
This arises from the fact that the sensitivity comes from the highest-energy event categories where the second order LV effects are the most important (see section~\ref{sec:LVPheno}). 
These results show that SK is extremely sensitive to LV. 
These sensitivities are respectively four and eight orders of magnitude better than the best limits on the isotropic coefficients $a$ and $c$ in the neutrino sector while comparable to the ones on the anisotropic coefficients~\cite{datatables}.

These results were obtained using the perturbative model that requires $|\delta h| \ll 1/L$, which we translated into $|\delta h| \leq 10\%/L$. In the sensitivity study for $a$ and $c$, it appeared that, respectively, 36.1\% and 1.7\% of the events used in the analysis did not satisfy this perturbative criterion. These events correspond to the longest distances and highest energies, which means that cutting them will result in a loss of sensitivity. Moreover, the distance in SK is not accurately known on an event by event basis while the energies above $\sim$10 GeV cannot be measured. It therefore appears that in contrast to beam experiments that have a fixed distance and given energy, the perturbative model is not suitable for SK atmospheric neutrinos. In the future, we intend to perform an improved analysis by using the full SME.
Furthermore, neutrino oscillation experiments using either the short-baseline approximation or the perturbative model to look for LV effects should keep in mind that the perturbative criterion does not allow the calculation of LV expectation above it. In such cases, experiments should report their results in terms of a band of LV values excluded rather than simple limits.


\begin{thebibliography}{x}

\bibitem{SME}
V.A.\ Kosteleck\'y and M.\ Mewes,
Phys.\ Rev.\ D {\bf 85}, 096005 (2012).

\bibitem{NuLV}
LSND Collaboration,
L.B.\ Auerbach \etal,
Phys.\ Rev.\ D {\bf 72}, 076004 (2005);
MINOS Collaboration,
P.\ Adamson \etal,
Phys.\ Rev.\ Lett.\ {\bf 101}, 151601 (2008);
Phys.\ Rev.\ Lett.\ {\bf 105}, 151601 (2010);
Phys.\ Rev.\ D {\bf 85}, 031101 (2012);
MiniBooNE Collaboration,
A.A.\ Aguilar-Arevalo \etal,
Phys.\ Lett.\ B {\bf 718} (2013);
IceCube Collaboration,
R.\ Abbasi \etal, 
Phys.\ Rev.\ D {\bf 82}, 112003 (2010).

\bibitem{SKNIM}
Super-Kamiokande Collaboration,
S.\ Fukuda \etal,
NIM A {\bf 501}, 418-462 (2003).

\bibitem{SK1998}
Super-Kamiokande Collaboration,
Y.\ Fukuda \etal,
Phys.\ Rev.\ Lett.\ {\bf 81}, 1562 (1998).

\bibitem{PertLV}
J.S.\ D\'\i az \etal,
Phys.\ Rev.\ D {\bf 80}, 076007 (2009).

\bibitem{SK3flavors}
Super-Kamiokande Collaboration,
R.\ Wendell \etal,
Phys.\ Rev.\ D {\bf 81}, 092004 (2010).

\bibitem{datatables}
{\it Data Tables for Lorentz and CPT Violation,}
V.A.\ Kosteleck\'y and N.\ Russell,
2013 edition,
arXiv:0801.0287v6.

\end{thebibliography}
\end{document}